\documentclass{ws-mpla}
\usepackage[super]{cite}
\usepackage{graphicx,graphbox,xcolor}

\renewcommand{\d}{\mathrm{d}}

\begin{document}

\markboth{M. E. A. Gadja, L. Khodja \& Y. Delenda}{Higgs $p_t$ distribution in $h+$jet production at hadron colliders within the NC-ESM}

\catchline{}{}{}{}{}

\title{Higgs $p_t$ distribution in Higgs $+$ jet production at hadron colliders within the noncommutative effective standard model}

\author{Mohamed El Arebi Gadja$^*$ and Lamine Khodja$^\dag$}

\address{Laboratoire de Rayonnement et Plasmas et Physique de Surfaces,\\
D\'{e}partement de Physique, Facult\'{e} des Math\'{e}matiques et des Sciences de la Mati\`{e}re,\\
Universit\'{e} Kasdi Merbah Ouargla, Ouargla 30000, Algeria\\
$^*$gadja.mohamed@univ-ouargla.dz\\
$^\dag$khodja.lamine@univ-ouargla.dz}

\author{Yazid Delenda$^\ddag$\footnotetext{$^\ddag$Corresponding author.}}

\address{Laboratoire de Physique des Rayonnements et de leurs Interactions avec la Mati\`{e}re,\\
D\'{e}partement de Physique, Facult\'{e} des Sciences de la Mati\`{e}re,\\
Universit\'{e} de Batna-1, Batna 05000, Algeria\\
yazid.delenda@univ-batna.dz}

\maketitle


\begin{abstract}
We use the noncommutative Higgs effective standard model to make a phenomenological prediction for the transverse momentum distribution of the Higgs boson produced in association with a jet at hadron colliders. We calculate at leading order in the noncommutative parameter $\Theta$ as well as leading order in the strong coupling $\alpha_s$, the one-loop $p_t$ distribution of the Higgs boson. As in the standard model, the fixed-order distribution suffers from large logarithms at small $p_t$ which require an all-orders resummation. We find that the large-$p_t$ region of the distribution is strongly affected by the non-commutativity, while small-$p_t$ region is not. Following this observation, we propose a simple matching method that allows us to compute a result that is also valid at small $p_t$ obtained with standard-model parton showers such as \texttt{Pythia 8}. We also compare our results with the next-to-leading order (NLO) and next-to-next-to-leading order (NNLO) distributions in the standard model, in order to assess the importance of higher-order effects in the search for non-commutativity at colliders.
\end{abstract}

\ccode{PACS Nos.: 11.10.Nx, 14.80.Bn}

\section{Introduction}	

Following the discovery of the Higgs boson in 2012 \cite{aad2012observation,chatrchyan2012observation}, the main focus of Higgs physics has entered the era of precision phenomenology by measuring its properties, specifically its kinematical behavior, couplings to Standard Model particles, and new physics searches \cite{buckley2021comparative}.  The Higgs production cross-section and decay rate are used to study the coupling between the Higgs boson and other particles. In this regard, the production of a high-$p_t$ jet alongside the Higgs boson is a process of great interest both in making precision measurements as well as the search for new-physics phenomenon at hadron colliders.

The correlation between the Higgs particle and the jet will undoubtedly provide important information and further reveal the electroweak coupling of the Higgs boson \cite{dawson1991radiative,djouadi1991production,de1999higgs, ravindran2002next}. In this work, we study the process of production of a Higgs plus jet at the LHC within the framework of the noncommutative standard model. We specifically consider an important observable used intensively to extract properties of the Higgs boson, namely the transverse momentum ($p_t$) distribution of the produced Higgs.

The Higgs field is somehow related to the noncommutative nature of space-time, an idea which emerged in the late 1980s and early 1990s \cite{connes1991particle, dubois1989classical}, and was included in the description of the Standard Model in Refs. \refcite{chamseddine1997spectral} and \refcite{connes1996gravity}. Several phenomenological works have been performed in the literature that aim at studying the applications of the non-commutativity of space-time in particle physics, such as Refs. \refcite{chaichian2003non,buric2007one,ohl2004testing,castro2015decays,haghighat2006photon,melic2005standard}.

More precisely, noncommutative geometry \cite{Alain1994} is a branch of mathematics developed by Alain Connes that aims to generalize geometric ideas to spaces where coordinates do not commute. Driven by quantum gravity, space-time is assumed to be a noncommutative manifold at very high energy scales. Even if the exact nature of this noncommutative manifold remains unknown, it seems reasonable to assume that at an intermediate scale, say several orders of magnitude lower than the Planck scale, the corresponding coordinate algebra is only a slightly noncommutative matrix algebra \cite{stephan2007almost}.

In the canonical version of the noncommutative space-time one has
\begin{equation}
[\hat{x}^\mu,\hat{x}^\nu]=i\,\Theta^{\mu\nu}\,,
\end{equation}
where $\Theta^{\mu\nu}$ is a real constant anti-symmetric tensor, and $\hat{x}$ denotes the noncommutative coordinates. The action in noncommutative field theories is obtained by using the Moyal-star product and the Seiberg-Witten maps \cite{seiberg1999string}, where the star product of two commutative fields $\psi$ and $\phi$ is defined by \cite{douglas2001noncommutative,riad2000noncommutative}
\begin{equation}
\psi(x)\star\phi(x)=\psi(x)\exp\left[\frac{i}{2}\,\overleftarrow{\partial}_\mu\,\Theta^{\mu\nu}\,\overrightarrow{\partial}_\nu\right]\phi(x)\,,
\end{equation}
and the ordinary spinor ($\psi$) and gauge ($V_\mu$) fields transform via the Seiberg-Witten maps as \cite{seiberg1999string,melic2005standard,uelker2008seiberg,ulker2012all}
\begin{subequations}
\begin{align}
\hat{\psi}(x,\Theta) &=\psi(x)+\Theta\,\psi^{(1)}[V]+\mathcal{O}(\Theta^2)\,,\\
\hat{V}_\mu(x,\Theta)&=V_\mu+\Theta\,V_\mu^{(1)}[V]+\mathcal{O}(\Theta^2)\,,
\end{align}
\end{subequations}
where to leading order in the $\Theta$ parameter\cite{Jurco:2001rq,Wohlgenannt:2003de,uelker2008seiberg}
\begin{subequations}
\begin{align}
\Theta\,\psi^{(1)}[V]&=-\frac{1}{2}\,\Theta^{\mu\nu}\,V_\mu\,\partial_\nu\psi+\frac{i}{4}\,\Theta^{\mu\nu}\,V_\mu\,V_\nu\,\psi\,,\\
\Theta\, V^{(1)}_\mu[V]&=-\frac{1}{4}\Theta^{\alpha\beta}\{V_\alpha,\partial_\beta V_\mu+F_{\beta\mu}\}\,,
\end{align}
\end{subequations}
where $F_{\beta\mu}$ is the QCD field strength tensor and $\{\cdots,\cdots\}$ stands for the anti-commutator.

In this paper, we use the noncommutative Higgs effective field theory (NC-HEFT) in order to make a phenomenological study of the Higgs transverse momentum distribution. We calculate within this framework the production cross-section of the Higgs boson, and estimate at leading order (LO) in the strong coupling the transverse momentum distribution. The distribution at fixed order suffers from large logarithms in the ratio $p_t/Q$, where $Q$ is the hard scale of the process. To obtain a reliable result at low $p_t$, where interesting physics are usually found, a resummation to all orders is necessary. As it turns out, the distribution at small values of $p_t$ is not affected by the non-commutativity of space-time, while the tail of the distribution (high-$p_t$ region) is strongly affected by it. We therefore propose a matching method that allows us to combine the two regions of $p_t$ in order to obtain a distribution that is valid in the entire spectrum of $p_t$, and is suitable for phenomenology. Furthermore, we compare the matched result with the next-to-leading order (NLO) distribution obtained with \texttt{MCFM} \cite{Campbell:2019dru} for the process under consideration\cite{Neumann:2018bsx}, and with the next-to-next-to-leading order (NNLO) result obtained from Ref. \refcite{Boughezal:2015dra}, and discuss the possibility of distinguishing noncommutative effects from NLO and NNLO corrections.

This paper is organised as follows. In Sec. 2, we present the HEFT in the noncommutative geometry formalism, and calculate the cross-section of production of a Higgs boson and a single jet at LO in $\alpha_s$ in $pp$ collisions at the LHC. We then, in Sec. 3, extract the Higgs $p_t$ distribution at LO. We use \texttt{MadGraph5\_aMC@NLO} \cite{Maltoni:2002qb, Alwall:2014hca} and \texttt{MadAnalysis5} \cite{Conte:2012fm} in order to perform the convolution with parton distribution functions (pdfs) and include experimental cuts. Furthermore, we use \texttt{Pythia8} \cite{Sjostrand:2014zea,Alwall:2008qv} parton shower  to produce a SM distribution, with which we match our NC distribution to obtain a result valid for all values of $p_t$, and compare this distribution with NLO and NNLO results. Finally, in the last section, we draw our conclusions.

\section{Higgs + Jet Production in the NC-HESM}

\begin{figure}[ht]
\centerline{\includegraphics[width=.9\textwidth]{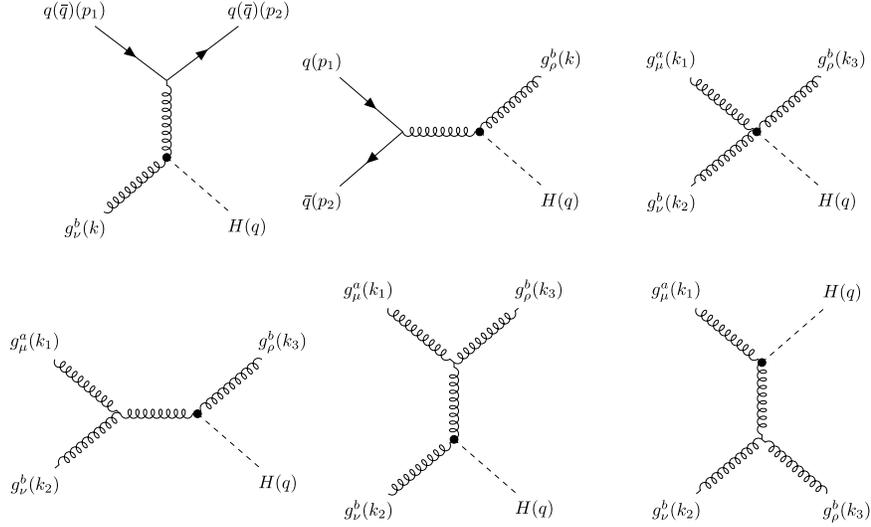}}
\vspace*{8pt}
\caption{Feynman diagrams contributing to the process $pp\to$ Higgs + jet. The dot represents the effective gluon-Higgs coupling in the infinite top-quark mass limit \cite{dawson1991radiative}. \protect\label{fig1}}
\end{figure}

As in the Higgs effective standard model (HESM), the  Higgs + jet production process in $pp$ collisions within the noncommutative Higss effective standard model (NC-HESM) has the
Feynman diagrams shown in Fig. \ref{fig1}.

\begin{figure}[ht]
\centerline{\includegraphics[align=c,width=0.27\textwidth]{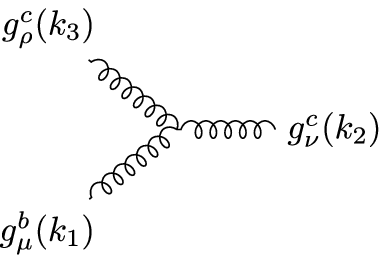}\qquad
$\displaystyle
\begin{aligned}
&g_s\,f^{abc}\left[g_{\mu\nu}(k_1-k_2)_\rho+g_{\nu\rho}(k_2-k_3)_\mu-g_{\rho\mu}(k_3-k_1)_\nu\right]\\
&\qquad+\frac{1}{2}\,g_s\,d^{abc}\,\Theta_{3\,\mu\nu\rho}
\end{aligned}
$}
\vskip 0.5cm
\centerline{\includegraphics[align=c,width=0.27\textwidth]{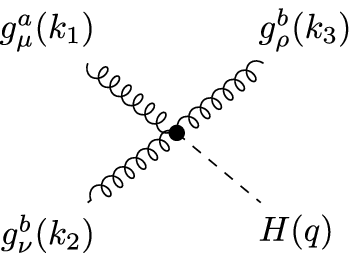}\qquad
$\displaystyle
\begin{aligned}&G_H\,g_s\,f^{abc}\left[g_{\mu\nu}(k_1-k_2)_\rho+g_{\nu\rho}(k_2-k_3)_\mu-g_{\rho\mu}(k_3-k_1)_\nu\right]\\
&\qquad+G_h\,g_s\,f^{abc}(k_1+k_2+k_3)^\alpha \epsilon_{\mu\nu\rho\sigma}\\
&\qquad-\frac{1}{2}G_H\,g_s d^{abc}\,\widetilde{\Theta}_{1\mu\nu\rho} -G_h\,g_s\,d^{abc}\,\widetilde{\Theta}_{2\,\mu\nu\rho}
\end{aligned}$}
\vspace*{8pt}\caption{Deformed vertices by noncommutative geometry. \protect\label{fig01}}
\end{figure}

The interaction point of two gluons with the Higgs boson in ordinary Higgs effective field theory has been calculated in Ref. \refcite{dedes2017feynman}. The noncommutative geometry at first order in $\Theta$ results solely in a correction term to the triple gluon vertex as well as the triple-gluon + Higgs boson vertex, as shown in Fig \ref{fig01}.
In this figure, $g_s$ is the strong coupling, $f_{abc}$ and $d_{abc}$ are, respectively, the SU(3) structure constants and $d$ symbols, $\epsilon$ is the totally anti-symmetric Levi-Civita tensor, and $k_i$ represent the momenta of the gluons. The Wilson coefficients are given by
\begin{subequations}
\begin{align}
G_H&=-\frac{g_s^2}{12\pi ^2v}\left(1+\frac{7\,m_H^2}{120\,m_t^2}+\frac{m_H^4}{168\,m_t^4}+\frac{13\,m_H^6}{16800\,m_t^6}\right),\\
G_h&=-\frac{g_s^2}{8\pi^2v}\left(1+\frac{m_H^2}{12\,m_t^2}+\frac{m_H^4}{90\,m_t^4}+\frac{m_H^6}{560\,m_t^6}\right),\label{eq 02}
\end{align}
\end{subequations}
with $m_t$ and $m_H$ being the masses of top quark and Higgs boson, respectively, and $v$ is Higgs vacuum expectation value. In this figure, $\Theta_3$ has been calculated in Ref. \refcite{melic2005standard} and is given by
\begin{align}
\Theta_{3}^{\mu\nu\rho}=\,&-(k_1\Theta k_2)\left[(k_1-k_2)^\rho g^{\mu\nu}+(k_2-k_3)^\mu g^{\nu\rho}+(k_3-k_1)^\nu g^{\rho\mu}\right]\notag\\
& -\Theta^{\mu\nu}\left[k_1^\rho(k_2\cdot k_3)-k_2^\rho(k_1 \cdot k_3)\right]-\Theta^{\nu\rho}\left[k_2^\mu(k_3 \cdot k_1)-k_3^\mu(k_2 \cdot k_1)\right]\notag\\
&-\Theta^{\rho\mu}\left[k_3^\nu(k_1\cdot k_2)-k_1^\nu(k_3\cdot k_2)\right]+(\Theta k_2)^\mu\left[g^{\nu\rho}k_3^2-k_3^\nu k_3^\rho\right]\notag\\
&+(\Theta k_3)^\mu\left[g^{\nu\rho}k_2^2-k_3^\nu k_2^\rho\right]+(\Theta k_3)^\nu\left[g^{\mu\rho}k_1^2-k_1^\mu k_1^\rho\right]\notag\\
& +(\Theta k_1)^\nu\left[g^{\mu\rho}k_3^2-k_3^\mu k_3^\rho\right]+(\Theta k_1)^\rho\left[g^{\mu\nu}k_2^2-k_2^\mu k_2^\nu\right]\notag\\
& +(\Theta k_2)^\rho\left[g^{\mu\nu}k_1^2-k_1^\mu k_1^\nu\right],
\end{align}
where  $(k_i\Theta k_j)=k_{i\mu}\Theta^{\mu\nu}k_{j\nu}$, $(\Theta k)^\mu=\Theta^{\mu\nu}k_{\nu}$, and $(\Theta k)^2=(\Theta k)^\nu(\Theta k)_\nu$. In our kinematics, all these products vanish except for $(\Theta k_3)^2=\Theta^2 k_t^2$. Furthermore, $\widetilde{\Theta}_1$ and $\widetilde{\Theta}_2$ are computed in this work, in our choice of noncommutative geometry parameter $\Theta$, to be
\begin{subequations}
\begin{align}
\widetilde{\Theta}_1^{\mu\nu\rho}
&=(\Theta k_3)^\nu k_1^\rho k_2^\mu+(\Theta k_3)^\mu k_1^\nu k_2^\rho+(\Theta k_3)^\nu k_1^\rho k_3^\mu\notag\\
&+(\Theta k_3)^\mu k_2^\rho k_3^\nu+\Theta^{\nu\rho}k_3^\mu\,k_1\cdot k_2+\Theta^{\mu\rho}k_3^\nu\,k_1\cdot k_2\notag\\
&-(\Theta k_3)^\nu g^{\mu\rho}\left(k_1\cdot k_2+k_1\cdot k_3\right)-(\Theta k_3)^\mu g^{\nu\rho}\left(k_1\cdot k_2+k_2\cdot k_3\right)\notag\\
&-\left(\Theta^{\mu\rho}k_1^\nu +\Theta^{\mu\nu}k_1^\rho\right) k_2\cdot k_3-\left(\Theta^{\nu\rho}k_2^\mu+\Theta^{\mu\nu}k_2^\rho\right)k_1\cdot k_3\,,\\
\widetilde{\Theta}_2^{\mu\nu\rho}
&=\Theta^{\nu\rho}\epsilon^{\mu\alpha\beta\gamma}k_{1\gamma} k_{2\beta} k_{3\alpha} +\Theta^{\mu\rho}\epsilon^{\nu\alpha\beta\gamma}k_{1\beta} k_{2\gamma} k_{3\alpha}\notag\\
&+\Theta^{\mu\nu}\epsilon^{\rho\alpha\beta\gamma}k_{1\beta} k_{2\alpha} k_{3\gamma}+2(\Theta k_3)^\nu \epsilon^{\mu\rho \alpha\beta}k_{1\alpha} k_{2\beta}\notag\\
&+2(\Theta k_3)^\mu\epsilon^{\nu\rho\alpha\beta}k_{1\alpha} k_{2\beta}\,.
\end{align}
\end{subequations}

In this paper we work with space-space non-commutativity, i.e., $\Theta^{i0}=0$, motivated by the unitary problem \cite{gomis2000space}, and without loss of generality we choose $\Theta^{21}=\Theta$ and $\Theta^{13}=\Theta^{23}=0$. Then, the color and polarization-averaged/summed squared amplitude of the last four Feynman diagrams in Fig. \ref{fig1} is found to be
\begin{align}
\overline{\left|\mathcal{M}^{\mathrm{NC}}_{gg\to hg}\right|^2}=\,&\overline{\left|\mathcal{M}^{\mathrm{SM}}_{gg\to hg}\right|^2}
+\Theta^2\,g_s^2\,\frac{\mathrm{C_A^2}-4}{128\,\mathrm{C_A^2C_F^2}}\,\Big[-\left(162G^2_h+47G^2_H\right)stu\notag\\
&+4G^2_h\,k_t^2 \left(-3 s m_H^2+48 s^2+47 s (t+u)+24 t^2-2t u+24 u^2\right)\notag\\
&+G^2_H\,k_t^2\left(5 s m_H^2+21 s^2+53 s t+57 s u+27 t^2-3t u+35 u^2\right)\Big]\,,
\end{align}
where $\mathrm{C_F}=4/3$ and $\mathrm{C_A}=3$ are the Casimir scalars in the fundamental and adjoint representations of SU(3), $s$, $t$, and $u$ are the usual Mandelstam variables, and $k_t$ is transverse momentum of the outgoing gluon. Here $\mathcal{M}^{\mathrm{SM}}_{gg\to hg}$ is the amplitude of the process $gg\to hg$ within the HESM. Additionally, the color/spin/polarization-averaged/summed squared amplitude of the first and second diagrams in our choice of $\Theta$ is not affected by non-commutativity, that is\footnote{Here the $q$ also implies processes with antiquarks.}
\begin{subequations}
\begin{align}
\overline{\left|\mathcal{M}^{\mathrm{NC}}_{gq\to hq}\right|^2}&=\overline{\left|\mathcal{M}^{\mathrm{SM}}_{gq\to hq}\right|^2}\,,\\
\overline{\left|\mathcal{M}^{\mathrm{NC}}_{q\bar{q}\to hg}\right|^2}&=\overline{\left|\mathcal{M}^{\mathrm{SM}}_{q\bar{q}\to hg}\right|^2}\,.
\end{align}
\end{subequations}
Then, the cross-section for the process $pp\to h+j$ reads
\begin{equation}
\sigma^{\mathrm{NC}}=\sum_{\delta}\int dx_adx_bf_a(x_a,\mu_\mathrm{F}^2)f_b(x_b,\mu_\mathrm{F}^2)\,\Xi\,\frac{\d p_t^2\d y}{16\pi^2s}\,\delta(s+t+u-m_H^2)\overline{\left|\mathcal{M}^{\mathrm{NC}}_{\delta}\right|^2}\,,
\end{equation}
where the sum extends over all the partonic channels $\delta$, $x_a$ and $x_b$ are the momentum fractions carried by the incoming partons, and $f_a$ are the pdfs evaluated at the factorization scale $\mu_{\mathrm{F}}$, and $\Xi$ denotes the experimental cuts. The $p_t$ and $y$ represent the transverse momentum and rapidity of the Higgs boson.

\section{Inclusive Cross-section and Transverse Momentum Distribution}

Having obtained the squared amplitude in NC geometry, we now proceed to calculate the total cross-section as well as the Higgs $p_t$ distribution. To do so, a straightforward method is to use Monte Carlo event generators. For this purpose, we use \texttt{MadGraph5\_aMC@NLO} \cite{Maltoni:2002qb, Alwall:2014hca} in order to generate a sample of parton-level events for the process $pp\to H j$ at LO within the SM-HEFT. To incorporate NC geometry all we have to do is modify the weight of each event by replacing the matrix element squared of the SM by the corresponding NC one. Said differently, we may write the total cross-section in NC geometry as
\begin{equation}\label{eq:wt}
\sigma_{\mathrm{NC}} = \frac{\sigma_{\mathrm{SM}}}{N}\sum_{\mathrm{events}} \left(\overline{\left|\mathcal{M}^{\mathrm{NC}}_{\delta}\right|^2}/ \overline{\left|\mathcal{M}^{\mathrm{SM}}_{\delta}\right|^2}\right),
\end{equation}
where the sum extends over all $N$ events in the generated sample, and $\sigma_{\mathrm{SM}}$ is the SM-HEFT cross-section obtained with \texttt{MadGraph5\_aMC@NLO}, and
$\delta$ is the corresponding channel. For each event record we access the particle kinematics, using \texttt{MadAnalysis5} \cite{Conte:2012fm}, and compute the corresponding weight as shown in Eq. \eqref{eq:wt}.

To present our quantitative results, we consider $pp$ scattering at $\sqrt{s}= 14 \,\mathrm{TeV}$ and fix the pole mass of the Higgs boson $m_H$ to the value $m_H=125\,\mathrm{GeV}$. We use NNPDF3.1 NNLO pdfs \cite{NNPDF:2017mvq} through \texttt{LHAPDF} \cite{Buckley:2014ana}, and impose the cut $\eta<5$ for the jet rapidity. To avoid the soft/collinear singularities in the inclusive cross-section, we cut the jet transverse momentum at 20 GeV. Further experimental cuts, for instance on the decay products of the Higgs boson, can also be applied within \texttt{MadGraph5\_aMC@NLO} when comparing to experimental data.\footnote{The Higgs is set not to decay in this work.} For our analysis, we set the renormalization and factorization scales at $\mu_\mathrm{F}=\mu_\mathrm{R}=m_H$. The SM cross-section that we obtain is
\begin{align}
\sigma^{\mathrm{SM}}=(13.110 \pm 0.001)\, \mathrm{pb}\,,
\end{align}
and the results in the NC-HESM are shown in Table \ref{ta1} for different values of the $\Theta$ noncommutative parameter.
\begin{table}[t]
\tbl{Cross-section for the production of the Higgs boson in association with a jet in NC geometry for different choices of the $\Theta$ parameter.}
{\begin{tabular}{@{}ccc@{}} \toprule
NC parameter $\Theta$ & & NC cross-section $\sigma$  \\
($\times 10^{-4}$ GeV$^{-2}$)&& $(\mathrm{NC})$ (pb) \\
\colrule
$0.1$ && $13.98$\\
$0.3$ &&$20.94$  \\
$0.5$ && $34.85$ \\
$0.7$ && $55.72$\\
 \botrule
\end{tabular}\label{ta1}}
\end{table}

\begin{figure}
	\centering
	\includegraphics[width=0.49\textwidth]{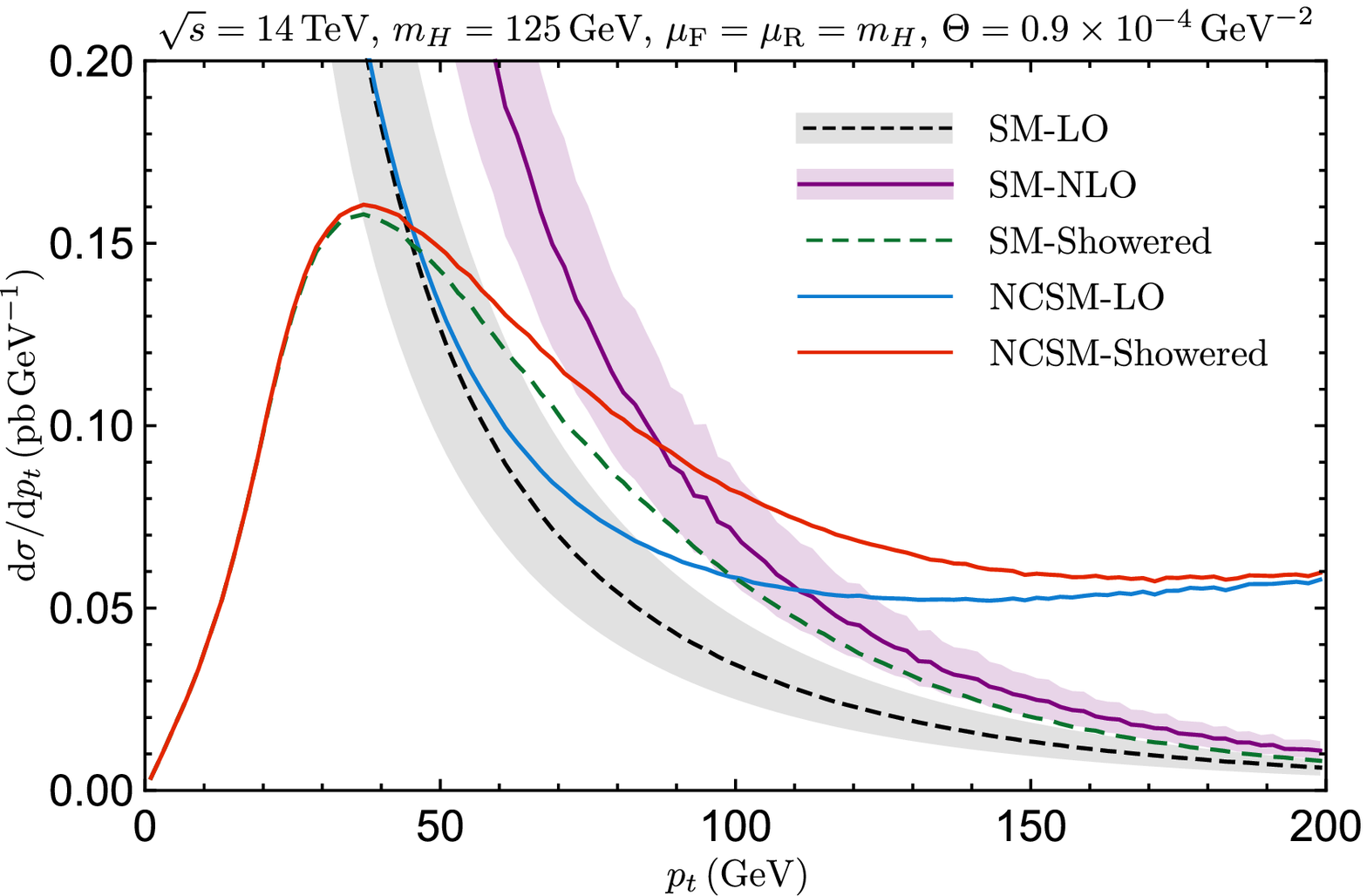}
    \includegraphics[width=0.49\textwidth]{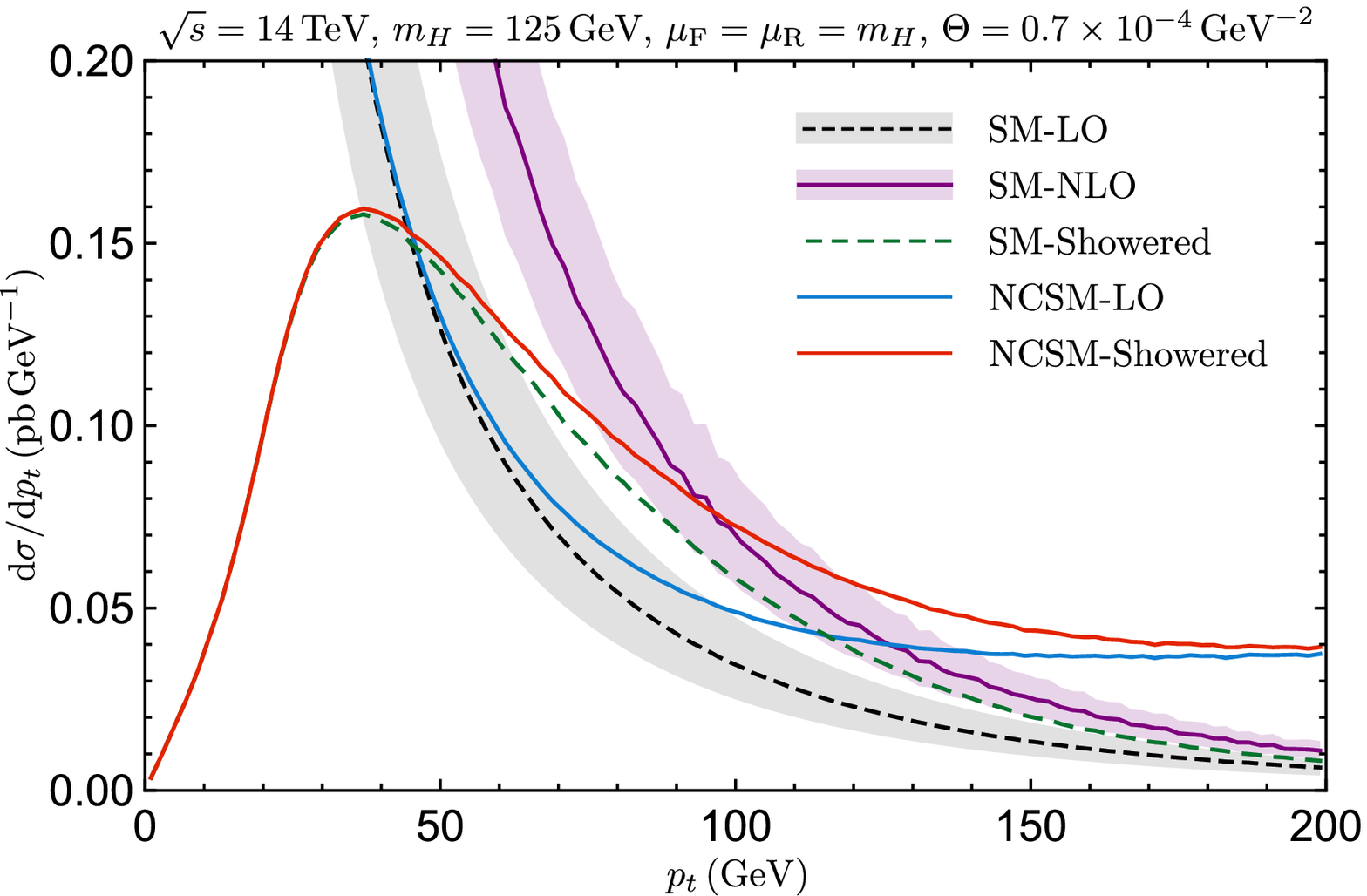}
    \includegraphics[width=0.49\textwidth]{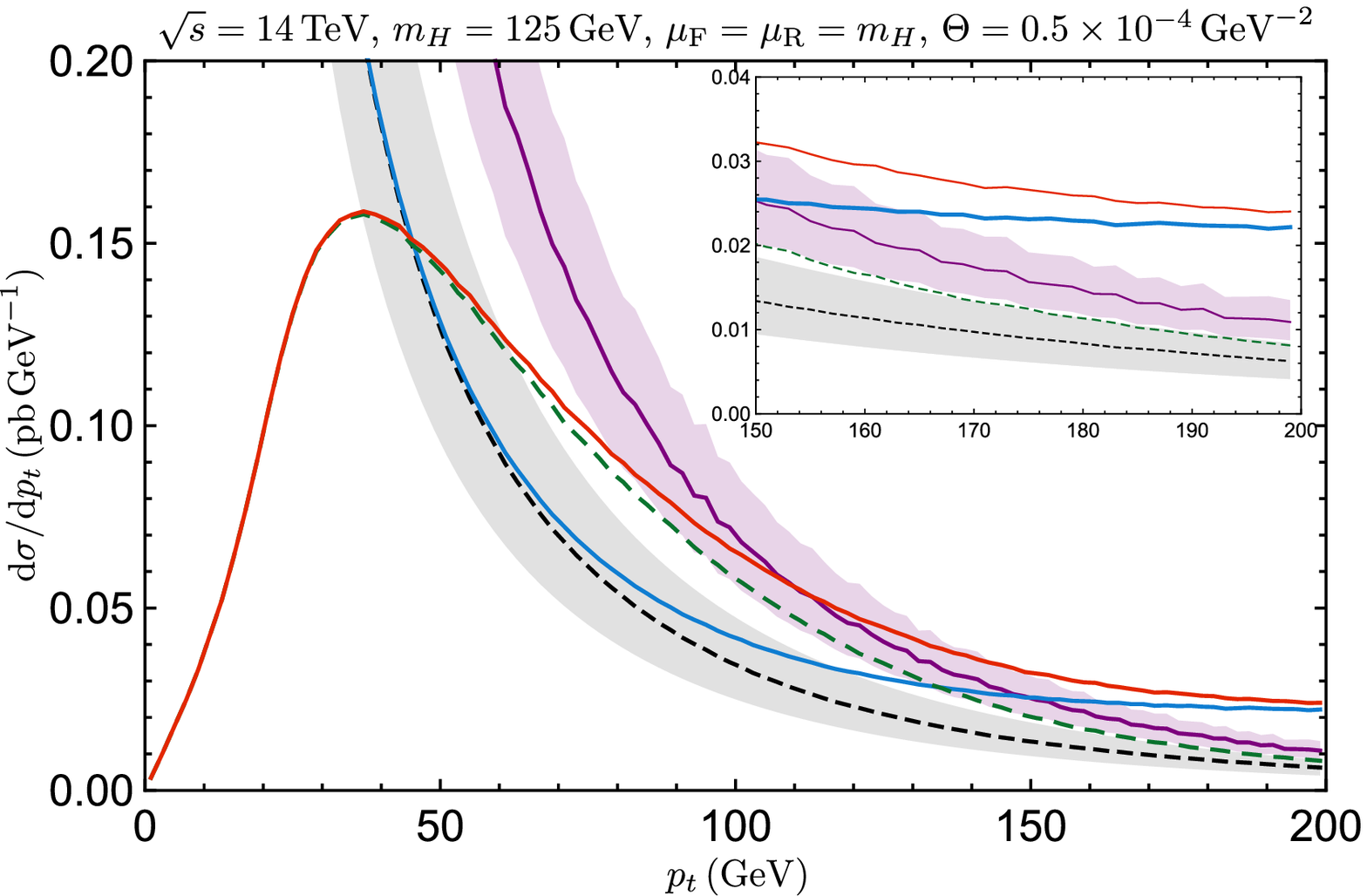}
    \includegraphics[width=0.49\textwidth]{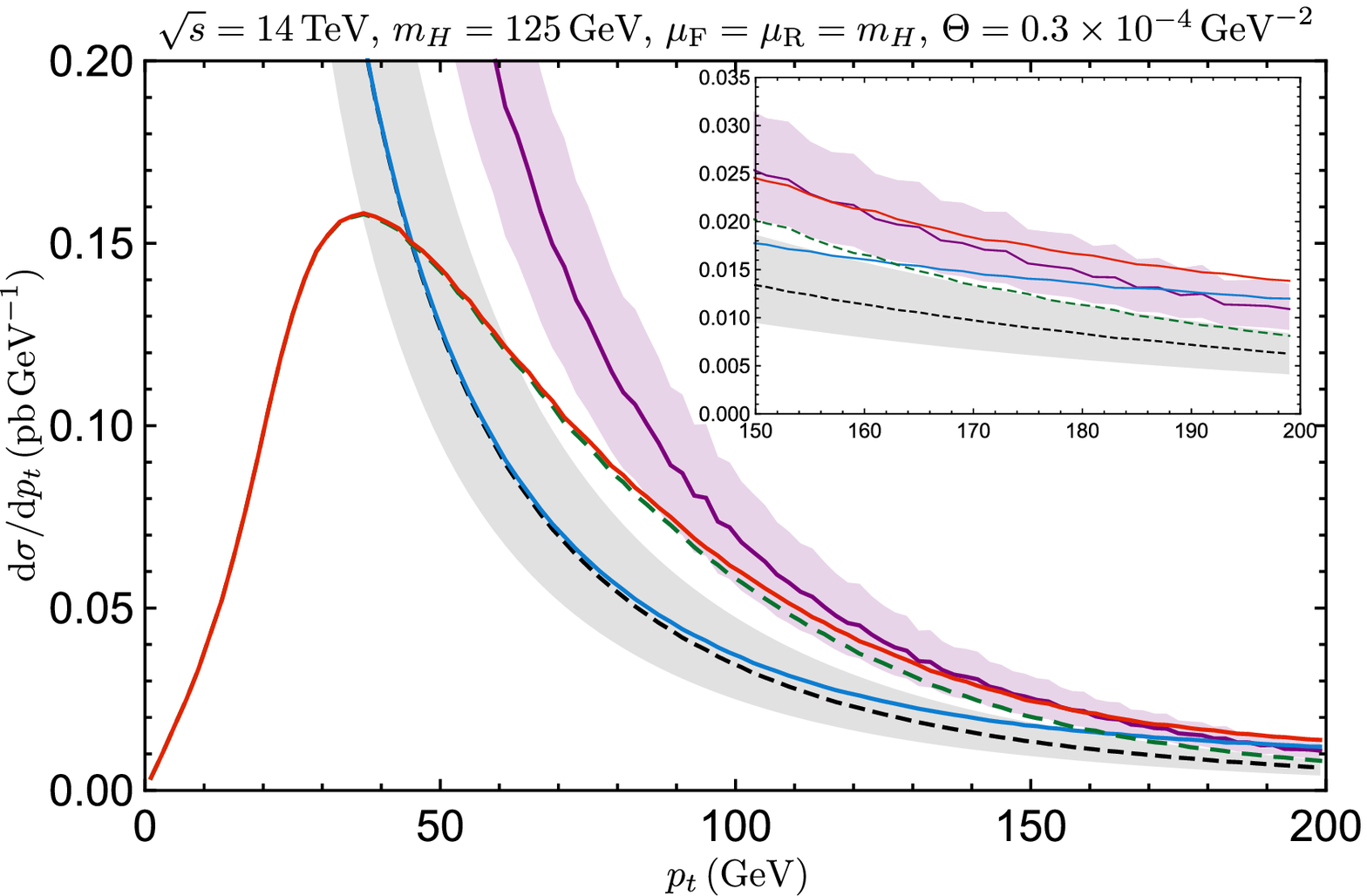}
	\vspace*{8pt}
	\caption{Higgs $p_t$ distribution. \protect\label{fig2} }
\end{figure}

We also calculate the Higgs $p_t$ distribution in a similar way. The results are shown in Fig. \ref{fig2} for different values of the non-commutativity parameter $\Theta$. As we can see from the curves, the fixed-order (LO and NLO) distribution is divergent for small values of $p_t$ both in the SM and NC-SM. This divergence has a leading double logarithmic structure that goes like $\alpha_s \ln^2(m_H^2/p_t^2)$ as well as sub-leading single logarithms $\alpha_s \ln(m_H^2/p_t^2)$. These logarithms are persistent at higher orders and may be resummed at next-to-leading logarithmic (NLL) accuracy into an exponential form
\begin{equation}
\sigma(p_t) \sim \exp\left( Lg_1(\alpha_sL)+g_2\alpha_s(L) \right),
\end{equation}
with $L=\ln (m_H^2/p_t^2)$ being the large logarithm, and $g_1$ and $g_2$ being resummation functions of $\alpha_sL$.

We notice that the $\mathcal{O}(\Theta^2)$ noncommutative correction to the squared matrix element does not manifest a soft/collinear divergence when $k_t\to0$. The source of the above-mentioned large logarithms is purely the SM contribution to the squared amplitude. For this reason, we can perform the resummation for the SM-like part of the distribution, and include the $\mathcal{O}(\Theta^2)$ corrections due to non-commutativity in the form of a matched distribution given by
\begin{equation}
\frac{\d\sigma_{\mathrm{matched}}^\mathrm{NC}}{\d p_t} = \frac{\d\sigma^\mathrm{SM}_{\mathrm{resummed}}}{\d p_t}-\frac{\d\sigma_\mathrm{LO}^\mathrm{SM}}{\d p_t}+ \frac{\d\sigma_\mathrm{LO}^\mathrm{NC}}{\d p_t}\,.
\end{equation}
This way, the expansion of the matched NC distribution at first order in the strong coupling exactly reproduces the NC LO distribution (with the correct $\mathcal{O}(\Theta^2)$ terms). This matched distribution also remains valid for all values of $p_t$ owing to the resummation of the large logarithms in the low-$p_t$ region in the SM part.

We do not explicitly perform the resummation in this paper. However, we can use Monte Carlo Parton Shower results to represent a resummed distribution and perform the matching between the fixed-order result and the parton shower distribution. The results for this are shown in Fig. \ref{fig2}. Here we observe the reasonable behaviour of the Higgs $p_t$ distribution at low $p_t$ form the SM parton shower \texttt{Pythia8}, and we see that the matched NC distribution coincides with this distribution in this region. At large values of $p_t$, the matched distribution approaches the LO distribution and has a noticeable deviation from the SM prediction. This indicates that the signature of noncommutative geometry may be hinted from the tail of this distribution.

It is therefore important to assess the impact of noncommutative geometry on the large-$p_t$ region of the distribution compared to other QCD effects such non-perturbative dynamics and higher-order contributions, namely NLO and NNLO corrections. Luckily, this region is not strongly affected by the former (non-perturbative effects), and is usually accurately estimated using just fixed-order calculations (up to NNLO). Thus, in this work, we use the fixed-order Monte Carlo program \texttt{MCFM} \cite{Campbell:2019dru} to compute the Higgs $p_t$ distribution at NLO within the SM-HEFT. Scale uncertainties are also estimated by varying the renormalization and factorization scales by factors of $2$ and $1/2$ around their central values. The results are shown in Fig. \ref{fig2}. We notice that NLO corrections at the tail of the distribution have comparable size to noncommutative corrections for values of $\Theta \sim 0.3\times 10^{-4}\,\mathrm{GeV}^{-2}$. For larger values, $\Theta\gtrsim 0.5\times 10^{-4}\,\mathrm{GeV}^{-2}$, the noncommutative corrections are noticeably higher than the NLO corrections with scale uncertainties.

\begin{figure}
	\centering
	\includegraphics[width=0.88\textwidth]{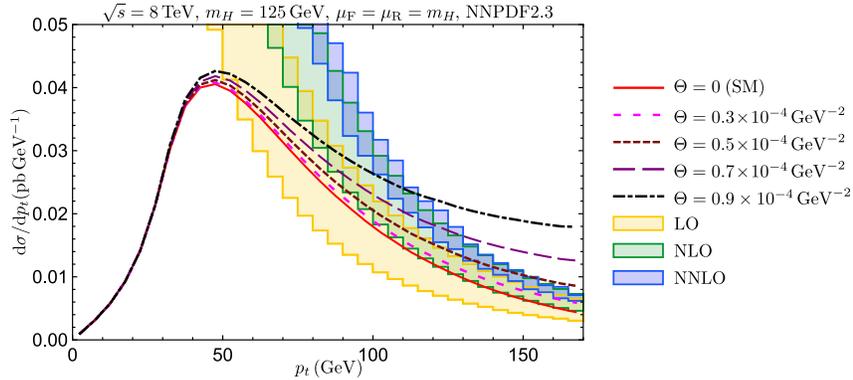}
	\vspace*{8pt}
	\caption{Comparison between Higgs $p_t$ distribution in the NC-SM and fixed-order results up to NNLO. The fixed-order results are extracted from Ref. \cite{Boughezal:2015dra}, and assume the infinite-top-mass limit. \protect\label{fig3} }
\end{figure}

Given the large size of NLO corrections, it is important to also consider the comparison of our results to the NNLO distribution. To achieve this, we use the result presented in Ref. \refcite{Boughezal:2015dra} for the NNLO Higgs $p_t$ distribution in which the $m_t\to\infty$ approximation is employed. The results are displayed in Fig. \ref{fig3} for $\sqrt{s}=8\,\mathrm{TeV}$, and where we use the NNPDF2.3 \cite{Ball:2011mu} pdf set and impose the cut on the jets $p_t>30$ GeV. We observe that the NNLO scale-uncertainty band almost entirely overlaps with the NLO one at the tail of the distribution, indicating that NNLO corrections are very small in this large-$p_t$ region. The scale uncertainties at NNLO are a lot smaller than those at NLO. This means that the values of $\Theta\gtrsim 0.5\times 10^{-4}\,\mathrm{GeV}^{-2}$ still lead to effects on the distribution that are clearly distinguishable from fixed-order corrections. This indicates that such large values of $\Theta$ can easily be confirmed or ruled out when compared to real experimental data of the same distribution. Additionally, an upper bound on the $\Theta$ parameter may be obtained from comparison to experimental data on this distribution --- an investigation that we leave to our forthcoming work.

\section{Conclusions}

In this work, we used the NC-HESM in order to perform detailed calculations that are relevant for Higgs phenomenology at the LHC. Specifically we calculated the noncommutative corrections to the SM squared amplitudes for the process of production of a Higgs boson in association with a jet at hadron colliders. We found that only the gluon-gluon fusion channel is affected by NC geometry at leading order in $\Theta$, while the other channels (which are actually suppressed in the SM) are unaffected.

Following this, we calculated the total cross-section for the production of the Higgs boson and the jet in the NC-HESM using \texttt{MadGraph5\_aMC@NLO} by modifying the weights of the events to include noncommutative corrections to the squared amplitudes. Furthermore, we calculated the differential distribution in the Higgs transverse momentum at leading order in the strong coupling. We also proposed a simple matching formula that allowed us to extend the range of validity of the distribution to low values of $p_t$, by showering (or resumming) the distribution in the SM and including the noncommutative corrections at leading order. Finally, we compared the size of noncommutative corrections for different values of $\Theta$ to those obtained in QCD at NLO using the Monte Carlo program \texttt{MCFM} and at NNLO using the results of Ref. \refcite{Boughezal:2015dra}, and discussed how it is possible to confirm or rule out the impact of noncommutative geometry when comparing to experimental data of this distribution.

\section*{Acknowledgments}

This work is supported by PRFU research project B00L02UN050120230003. The authors wish to thank the Algerian Ministry of Higher Education and Scientific Research and DGRSDT for
financial support.

Some of the numerical calculations presented here have been performed in the HPC cluster at the University of Batna 2.

\bibliographystyle{ws-mpla}
\bibliography{ref}

\end{document}